\documentclass{elsart}
\usepackage{natbib}
\usepackage{amssymb}
\usepackage{graphicx}
\newcommand{\Btaunu} {\ensuremath{B^- \rightarrow {\tau}^{-} \bar{\nu}}}
\newcommand{\BRBtaunu} {\ensuremath{B(\Btaunu)}}
\begin{document}
\begin{frontmatter}

\title{Including gaussian uncertainty on the background estimate for
upper limit calculations using Poissonian sampling}

\author{Luca Lista}

\address{INFN Sezione di Napoli\\
Complesso Universitario di Monte Sant'Angelo, edificio G, I-80126, Napoli, Italy\\
}

%=============================================================================
% Insert here the abstract
%=============================================================================

\begin{abstract}

A procedure to include the uncertainty on the background estimate for
upper limit calculations using Poissonian sampling is presented for the case
where a Gaussian assumption on the uncertainty can be made. 
Under that hypothesis an analytic expression of the likelihood is 
derived which can be written in terms of polynomials defined 
by recursion. This expression may lead to a significant
speed up of computing applications that extract the upper limits
using Toy Monte Carlo.

\end{abstract}


\end{frontmatter}

%=============================================================================
% The document starts here:
% You can divide it in section, subsection, subsubsection, subparagraph
%=============================================================================

%============================================================================

\section{Introduction}

In searches for rare processes where
there is no statistical evidence of
the signal, it is often convenient to combine the results of
more independent selection channels in order to obtain 
the upper limit to the number of expected signal events.
A statistical procedure using a likelihood ratio approach
has been adopted to combine the results of the Higgs 
search done by the four LEP experiments~\cite{ref:lep1}~\cite{ref:lep2}.
The likelihood ratio estimator $Q$ is defined as:
\begin{equation}
  Q=\frac{{L}(s+b)}{{L}(b)}
\end{equation}
where ${L}(s+b)$ and ${L}(b)$ are the likelihood functions
in the hypotheses of the signal plus background and background 
only respectively.

In the case of a Poissonian sampling, only the number 
of events passing a number of independent selection channels
is used as observable to discriminate the hypotheses of signal
plus background versus background only.
The likelihood functions ${L}(s+b)$
and ${L}(b)$ can be written, for the Poissonian sampling,
as the product of Poisson probabilities:
\begin{eqnarray}
  {L}(s+b) & = &
  \prod_{i=1}^{n_{ch}}\frac{e^{-(s_i+b_i)}(s_i+b_i)^{n_i}}{n_i!} 
  \label{eq:lsb}
  \\
  {L}(b) & = &
  \prod_{i=1}^{n_{ch}}\frac{e^{-b_i}b_i^{n_i}}{n_i!} 
  \label{eq:lb}
\end{eqnarray}
where $n_{ch}$ is the number of selection channels, $s_i$ and $b_i$
are the expected number of signal and background
events respectively and $n_i$ is the number of selected events.
The following simplified expression for $Q$ is more convenient
for computer computations:
\begin{equation}
  Q = \frac{{L}(s+b)}{{L}(b)} = e^{-s_{tot}}\prod_{i=1}^{n_{ch}}\left( 1 + \frac{s_i}{b_i} \right)^{n_i}
\,.
\end{equation}
where $s_{tot} = \sum_{i=1}^{n_{ch}} s_i$.

In the case where $s$ is the number of produced events and
$\epsilon_i$ are the efficiencies of each selection 
$s_{tot} = s \sum_{i=1}^{n_{ch}}\epsilon_i$.
The absolute minimum of $-2\log(Q)$ as a function of $s$ gives
the most likely value of the branching fraction.
In case there is no evidence of the signal,
it is possible to compute an upper limit at a given (usually
90\% or 95\%)
Confidence Level (C.L.) using a Toy Monte Carlo,
generating a large number of random experiments 
for different values of the
signal $s$. The confidence level for the signal
hypothesis can be computed as:
\begin{equation}
  C.L._s = \frac{C.L._{s+b}}{C.L._b} = 
  \frac{N_{Q_{s+b}\le Q}}{N_{Q_b\le Q}}
\end{equation}
where $N_{Q_{s+b}\le Q}$ and $N_{Q_b\le Q}$ are the number of
the generated experiments which have a likelihood ratio less then or equal
to the measured one, in the background plus signal and background only
hypothesis respectively. 

Background uncertainty can be included in the definition
of the likelihood applying a convolution with the
distribution of the background, which is given by the assumed
distribution of the background fluctuation.
The case where the error on the background estimates $b_i$ can 
be assumed to be Poissonian has been studied in ref.~\cite{ref:clpois}. 
In other cases, the uncertainty
may be better considered as Gaussian.
This is true for instance when
the background estimates are computed applying
subtractions of different samples or 
applying scaling factors affected by 
gaussian uncertainties.

It should be remarked that the method applied 
in reference~\cite{ref:clpois}, which is also applied in this paper,
does not have as good formal basis as it was originally thought.
Reference~\cite{ref:zech} notes that the method
is not correct from a frequentist point of view,
nonetheless, it ``seems to be acceptable to many pragmatic 
frequentists''. This caveat should be kept in mind when
handling the results reported in the following.

\section{Including Gaussian background uncertainties 
in the upper limit extraction}

Under the assumption that the uncertainties are Gaussian, equations
(\ref{eq:lsb}) and (\ref{eq:lb}) need to be convolved with
a Gaussian function, and become: 
\begin{eqnarray}
  {L}(s+b) & = &
  \prod_{i=1}^{n_{ch}}\int_{-\infty}^{+\infty}{\mathrm d}b_i^\prime 
  \frac{1}{\sqrt{2\pi \sigma_i^2}}e^{-\frac{(b_i^\prime - b_i)^2}{2\sigma_i^2}}
  \frac{e^{-(s_i+b_i^\prime)}(s_i+b_i^\prime)^{n_i}}{n_i!} 
  \\
  {L}(b) & = &
  \prod_{i=1}^{n_{ch}}
  \int_{-\infty}^{+\infty}{\mathrm d}b_i^\prime 
  \frac{1}{\sqrt{2\pi \sigma_i^2}}e^{-\frac{(b_i^\prime -
      b_i)^2}{2\sigma_i^2}}
  \frac{e^{-b_i^\prime}(b_i^{\prime})^{n_i}}{n_i!} 
\end{eqnarray}
where $\sigma_i$ is the error on the estimate of the background $b_i$.
The integration can be extended from $-\infty$ to $+\infty$
including the unphysical negative signal region when the area of the
tails of the Gaussian distributions in that region are negligible.
This is true if $\sigma_i$ is sufficiently smaller than $b_i$. 
In that case, the integrals are easily manageable analytically.
The integration can be performed exploiting the following expression for
the exponent of the exponential term:
\begin{equation}
  -\frac{(b_i^\prime - b_i)^2}{2\sigma_i^2} -(s_i+b_i^\prime) = 
  -\frac{(b_i^\prime- b_i + \sigma_i^2)^2}{2\sigma_i^2} - (s_i+b_i) + \frac{\sigma_i^2}{2}
\end{equation}
The intagration variable can be normalized to $z=(b_i^{\prime}- b_i +
\sigma_i^2)/\sigma_i $
so that the likelihood functions become:
\begin{eqnarray}
  {L}(s+b) & = &
  \prod_{i=1}^{n_{ch}}
  \frac{1}{\sqrt{2\pi}}
  \frac{1}{n_i !}e^{-(s_i+b_i)}e^{\sigma_i^2/2}
  \int_{-\infty}^{+\infty} e^{-z^2/2}
  (s_i+b_i-\sigma^2_i+\sigma_i z)^{n_i}{\mathrm d}z
  \\
  {L}(b) & = &
  \prod_{i=1}^{n_{ch}}
  \frac{1}{\sqrt{2\pi}}
  \frac{1}{n_i !}e^{-b_i}e^{\sigma_i^2/2}
  \int_{-\infty}^{+\infty} e^{-z^2/2}
  (b_i-\sigma^2_i+\sigma_i z)^{n_i}{\mathrm d}z
\end{eqnarray}
The integrals present in the likelihood functions are all of the form:
\begin{equation}
  \int_{-\infty}^{+\infty} e^{-z^2/2}
  (\alpha+\beta z)^n{\mathrm d}z
\end{equation}
and can be computed expanding the polynomial and using:
\begin{eqnarray}
  \int_{-\infty}^{+\infty} e^{-z^2/2}z^{2n}{\mathrm d}z & = &
  \sqrt{2\pi} \prod_{k=0}^n (2k+1) \\
  \int_{-\infty}^{+\infty} e^{-z^2/2}z^{2n+1}{\mathrm d}z & = & 0
\end{eqnarray}
or can be alternatively computed defining the $n$-th degree polynomials:
\begin{equation}
  p_n(\alpha, \beta) = 
  \frac{1}{\sqrt{2\pi}}\int_{-\infty}^{+\infty} e^{-z^2/2}
  (\alpha+\beta z)^n{\mathrm d}z\,.
\label{eq:poly}
\end{equation}
It can be easily shown that the polynomials 
satisfy the recursion relation:
\begin{equation}
  p_n(\alpha,\beta) = \alpha p_{n-1}(\alpha,\beta) + (n-1)\beta^2p_{n-2}(\alpha,\beta)\, .
\end{equation}
The first polynomials are:
\begin{eqnarray*}
  p_0(\alpha, \beta) &=& 1 \\
  p_1(\alpha, \beta) &=& \alpha \\
  p_2(\alpha, \beta) &=& \alpha^2+\beta^2 \\
  p_3(\alpha, \beta) &=& \alpha^3+3\alpha\beta^2 \\
  p_4(\alpha, \beta) &=& \alpha^4+6\alpha^2\beta^2+3\beta^4 \\
  p_5(\alpha, \beta) &=& \alpha^5+10\alpha^3\beta^2+15\alpha\beta^4 \\
  . . . & &
\end{eqnarray*}
The likelihood functions can be rewritten as:
\begin{eqnarray}
  {L}(s+b) & = &
  \prod_{i=1}^{n_{ch}}
  e^{-(s_i+b_i-\sigma_i^2/2)}
  \frac{ p_{n_i}(s_i+b_i-\sigma^2_i, \sigma_i)}{n_i !}\, ,
  \\
  {L}(b) & = &
  \prod_{i=1}^{n_{ch}}
  e^{-( b_i - \sigma_i^2/2)}
  \frac{p_{n_i}(b_i-\sigma^2_i, \sigma_i)}{n_i !}\, .
\end{eqnarray}
The computation of the upper limit with the Toy Monte Carlo can be
performed with a small modification to the code used for the
computation in the case of no error.
The following simplified expression for $Q$ may be convenient:
\begin{equation}
  Q = \frac{{L}(s+b)}{{L}(b)} = e^{-s_{tot}}\prod_{i=1}^{n_{ch}}\frac{p_{n_i}(s_i+b_i-\sigma_i^2,\sigma_i)}{p_{n_i}(b_i-\sigma_i^2, \sigma_i)}
\,.
\end{equation}

The usage of polynomials instead of other means of
numerical integration provides a significant sepeed up
of the code in computer calculations that use Toy Monte Carlo.

\section{Application to the search for \Btaunu\ in BABAR}

\begin{figure}[!htb]
  \begin{center}
    \includegraphics[height=6cm]{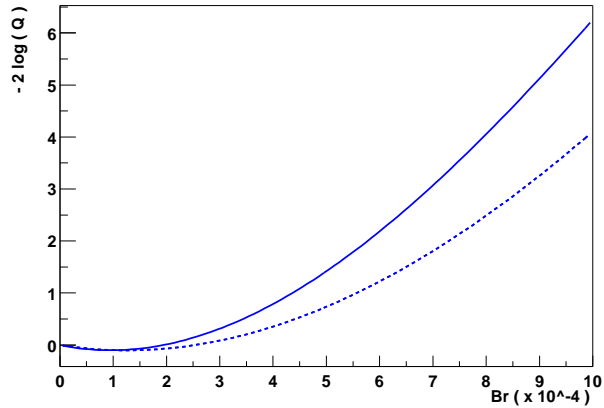}
    \caption{$-2\log(Q)$ as a function of \BRBtaunu\
      with (dashed) and without (solid) including the systematic error on the estimated background.}
    \label{fig:cl1}
  \end{center}
\end{figure}

\begin{figure}[!htb]
  \begin{center}
    \includegraphics[height=6cm]{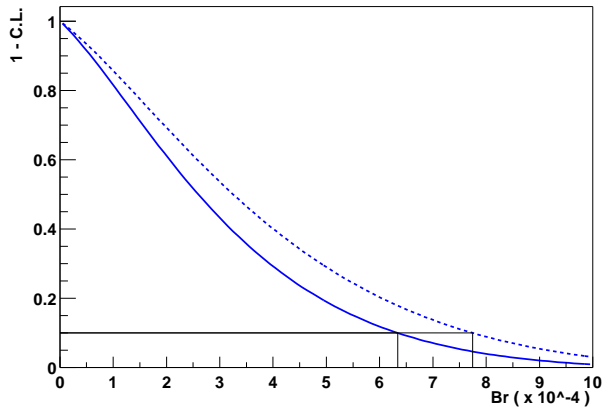}
    \caption{Confidence level as a function of \BRBtaunu\
      with (dashed) and without (solid) including the systematic error on the estimated background.}
    \label{fig:cl2}
  \end{center}
\end{figure}

The above method has been applyied in the search
for \Btaunu\ using in BABAR experiment~\cite{ref:babar}
using the events where one $B$ meson is fully reconstructed,
where the assumptions made in this paper hold. The
search has shown no evidence for the signal. 
In order to evaluate the 90\% C.L. limit on 
\BRBtaunu\ , a large number of Toy Monte Carlo experiments have been
generated for different values of the assumed branching fraction,
each corresponding to an expected number of produced events $s$. 
Fig.~\ref{fig:cl1} and.~\ref{fig:cl2} show 
$-2 \log(Q)$
and confidence level as a function
of \BRBtaunu\ with and without including the 
systematic error on the estimated background. 
The limit obtained including the uncertainty is
less stringent than the limit obtained neglecting
this effect.
\section{Conclusion}

A procedure to include Gaussian uncertainty on the background estimate for
upper limit calculations using Poissonian sampling 
has been presented. The likelihood can be written in terms of
polynominals that can be defined recursively. This approach
makes computing calculation of confidence levels
more efficient. The technique described in this paper has been applied 
in BABAR for to the search for \Btaunu\ .

%============================================================================

\begin{thebibliography}{9}

\bibitem{ref:lep1}
  The LEP Working Group for Higgs Boson Searches, 
  {\em Lower bound for Standard Model Higgs boson mass from combining
    the results of the four LEP experiments}, CERN-EP/98-046.
\bibitem{ref:lep2}
  The LEP Working Group for Higgs Boson Searches, 
  {\em Limits for the Higgs boson masses from combining the data of the
    four LEP experiments at $\sqrt{s}\le 183 GeV$}, CERN-EP/99-060.
  
\bibitem{ref:clpois}
  K.K.Gan {\em et al.}, {\em Incorporation of the statistical uncertainty
  in the background estimate into the upper limit on the signal}, 
  N.I.M. A412 (1998) 475-482.

\bibitem{ref:zech}
  G.Zech, {\em Frequentistic and Bayesian confidence limits},
  Eur.Phys.J.direct C4:12,2002. 

\bibitem{ref:babar}
  B. Aubert, et al., BABAR Collaboration,
  {\em  A Search for $B^- \rightarrow \tau^- \bar{\nu}$ Recoiling Against a Fully
  Reconstructed $B$}
  presented to 2003 Rencontres de
  Moriond: Electroweak Interactions and Unified Theories	
  hep-ex/0304030, BABAR-CONF-03/004, SLAC-PUB-9716

\end{thebibliography}
\end{document}